# Magnetoimpedance based detection of L-band electron paramagnetic resonance in 2,2-diphenyl-1-picrylhydrazyl (DPPH)

Ushnish Chaudhuri and R. Mahendiran [1]

Department of Physics, 2 Science Drive 3, National University of Singapore,

Singapore-117551, Republic of Singapore

Detection of electron paramagnetic resonance (EPR) using a microwave cavity resonating at a fixed frequency (between 9 and 10 GHz) remains the most popular method until now. Here, we report a cavity-less technique which makes use of only an impedance analyzer and a copper strip coil to detect L-band EPR ($f$ = 1-3 GHz) in the standard EPR marker 2,2-diphenyl-1-picrylhydrazyl (DPPH). Our method relies on measuring the magnetoimpedance (MI) response of DPPH through a copper strip coil that encloses DPPH. In contrast to commercial EPR which measures only the field derivative of power absorption, our method enables us to deduce both absorption and dispersion. Changes in resistance ($R$) and reactance ($X$) of the copper strip while sweeping an external dc magnetic field, were measured for different frequencies ($f$ = 0.9 to 2.5 GHz) of radio frequency current in the coil. $R$ exhibits a sharp peak at a critical value of the dc magnetic field, which is identified as the resonance field and $X$ shows a dispersion at the same frequency. The data were analyzed to obtain line width and resonance field parameters. The resonance field increased linearly with frequency and the obtained Landé $g$ factor of 1.999 is close to the accepted value of 2.0036, measured in *X*-band. The simplicity of this technique can be exploited to study paramagnetic centers in catalysis and other materials.

---

[1] Author for correspondence (phyrm@nus.edu.sg).

**Introduction**

The first electron paramagnetic resonance (EPR) spectra was recorded by Yevgeny Konstantovich Zavoisky almost 70 years ago using a simple home-built spectrometer[1]. Over the years, EPR has evolved into a sophisticated instrument and become an essential tool to detect unpaired electrons in solids which in turn provides information about electronic structure of the paramagnetic centers and their chemical environments.[2] EPR is also to probe local defects in $Si/SiO_2$,[3] to study structural phase transitions[4], kinetics of chemical reactions[5,6], electron and spin transfer in catalysis[7,8] structure of proteins, organic free radicals, etc.[7] Thus, applications of EPR span all the branches of science. Until now, the most popular method to detect EPR makes use of a fixed frequency microwave cavity resonator in the $X$ band (~9 GHz) or $Q$ band (~35 GHz) frequency regime[9]. However, metallic samples with high conductive losses as well as aqueous biological/chemical samples with large dielectric losses prove to be limiting for commercial EPR spectrometers operating in $X$ band since they cause frequency shift due to dispersion of permeability[10]. Recently, coplanar waveguides (CPW)[11,12,13,14] and microstripline resonators (MSR) [15,16,17,18] have been exploited to investigate smaller sized lossy samples over a broad frequency range. These methods take advantage of advances in microwave synthesizers working in broad frequency range (~50 MHz to ~50 GHz) to deliver microwave power to CPW/MSR to create intense local microwave magnetic field on the surfaces of CPW or MSR upon which a paramagnetic metallic sample is placed. However, with increasing frequency of the microwave signal, penetration of electromagnetic field inside a conducting sample also decreases due to skin effect. An EPR spectrometer working L-band (1- 3 GHz) or still lower frequencies is preferable to ensure a deeper penetration of electromagnetic waves into a conducting or a biological sample.

The EPR spectra at low frequencies can exhibit better resolution in certain cases, e.g. the EPR spectra of $Cu^{2+}$ complexes such as $Cu(DOPA)_2$ and $Cu(carnosine)_4$ were better resolved at 2.62 GHz than at 9.30 GHz [19]. Eaton and Eaton have given an overview of spectrometers developed for frequencies below X-band.[20] In this paper we describe a previously unreported technique for free radical compounds which can aid researchers investigate EPR from a fresh perspective.

In this article, we present a simple method to detect EPR in the L-band frequency region, which makes use of only an impedance analyzer and a copper stripcoil. Our method is compact, fast, requires very less instrumentation and can easily be incorporated in teaching and research laboratories. Our technique provides additional information pertaining to the absorptive and dispersive components of the high frequency magnetic susceptibility whereas conventional EPR spectrometers are designed to provide information about the field derivative of the power absorbed by the sample. Using our simple setup, we demonstrate the detection of EPR due to free radicals in a standard sample of 2,2-diphenyl-1-picrylhydrazyl (DPPH). As a stable and well-characterized solid radical source, DPPH is the most popular reference sample with Landé g-factor of 2.0036 [21]. The intensity of EPR signals depends on the number of radicals for a freshly prepared sample and can be determined by weighing the DPPH sample. DPPH exhibits a single response line in X-band with a small linewidth ~1.5–4.7 Oe due to the presence of only one unpaired spin per 41 atoms.

**Experimental Details**

To detect EPR in DPPH sample we measure the magnetoimpedance (MI) of a copper strip surrounding the sample. MI refers to the variation of electrical impedance ($Z\ (f,\ H) = R(f,H)+iX(f,H)$) of a material in presence of an applied *dc* magnetic field ($H_{dc}$) at different frequencies of alternating current (*f*). It consists of measuring the magnetic field dependence of the magnitude of impedance ($Z$) alone or resistance ($R$) and reactance ($X$) of the sample i.e., magnetoresistance and magnetoreactance, respectively.[22] Recently, we have discovered that MI measured by passing microwave (MW) current directly through some electrically conducting Mn-based perovskite oxides can detect paramagnetic as well as ferromagnetic resonances of exchange coupled $t_{2g}$ spins of $Mn^{3+}$:$t_{2g}^3e_g^1$ and $Mn^{4+}$:$t_{2g}^3e_g^0$ ions.[23,24] Since DPPH is insulating, high frequency current can't be injected through the sample and hence an indirect method is employed in the present work to measure the MI of DPPH. Our technique involves using a copper stripcoil as a detector and a radio frequency impedance analyzer (Agilent model E4991A) as a microwave signal source. DPPH powder obtained from Sigma-Aldrich$^{TM}$ was pressed into a disc shaped pellet at room temperature using a hydraulic press (pressure 5 ton/in$^2$). Then, the disc was cut into a rectangular bar of dimension (4.5mm x 3.5mm x 0.5mm). A 0.2 mm thick copper strip was folded in the shape of a cuboidal coil of the same dimension as that of the sample. The sample was tightly fixed inside the coil whose inner surface was covered with a Kapton tape to electrically insulate the sample from the copper strip. One end of the copper stripcoil was soldered to the signal line while the other end was soldered to the ground of a subminiature A type (SMA) coaxial connector. The radio frequency (*rf*) current from the impedance analyzer flows through the stripcoil and terminates at the ground of the SMA connector creating an *rf* magnetic field in the interior of the stripcoil along the axial direction as shown in Fig.1. Hence, the DPPH sample experiences an *rf* axial magnetic field. An electromagnet is used to apply *dc* magnetic field perpendicular to the axial *rf* field. The resistance (*R*) and reactive (*X*) components of the electrical impedance of the copper strip were simultaneous measured at different frequencies of *rf* current while sweeping the *dc* magnetic field. The electrical impedance of the copper strip is $Z = \frac{V_\phi}{I} = -\frac{1}{I}\frac{d\phi}{dt} = R + iX$, where $\phi$ is the *rf* magnetic flux passing through the stripcoil given by $\phi = \mu_0\mu_r H_{rf} A$. Here, $H_{rf}$ is the magnetic field inside the strip-coil and *A* is the cross-sectional area of the stripcoil. Since the high frequency permeability is $\mu = \mu' - i\mu''$ where $\mu'$ is the in-phase and $\mu''$ is the out-of phase of the permeability which describe dispersion and absorption or loss in the sample, respectively. By substituting the complex permeability, we obtain $R = G(\omega\mu_0\mu_r'')$ and $X = G(\omega\mu_0\mu_r')$, where G is a constant depending on the geometry of the stripcoil. Since the high frequency permeability of the paramagnetic DPPH sample is affected by the application of dc bias magnetic field, the resistance and reactance of the stripcoil also changes. *R* and *X* were recorded without and with the sample inside the stripcoil and the data for each frequency and magnetic field were subtracted to obtain only the sample contribution.

**Results and Discussions**

Fig. 1 (b) shows the magnetic field dependence of *R* and *X* for an *rf* current excitation of frequency 2 GHz in the stripcoil. As $H_{dc}$ is swept from $H_{dc} = 1$ kOe to $H_{dc} = 0$ Oe, *R* rapidly increases in a narrow field range and exhibits a sharp peak at 712 Oe whereas *X* shows a sudden jump around the same field. These features reflect the absorption and dispersion of the complex susceptibility, i.e., $\chi''$ and $\chi'$, respectively, in the vicinity of electron spin resonance[25]. Conventional EPR spectroscopy with lock-in detection technique measures the field derivative of the *rf* power absorbed, *dP/dH,* which is proportional to *dχ″/dH*. The dispersive signal is rarely reported except in a few experiments[26,27,28] even though it offers a better understanding of the spin dynamics in a material. The *R* and *X* responses were fitted to Eq. 1 and are shown as solid lines in Fig. 1 (b) which contains both a symmetric Lorentzian term (first component) and a dispersive antisymmetric term (second component).

$$R \text{ or } X = K_{sym}\frac{(\Delta H)^2}{(H_{DC} - H_r)^2 + (\Delta H)^2} + K_{asym}\frac{(\Delta H)(H_{DC} - H_r)}{(H_{DC} - H_r)^2 + (\Delta H)^2} + C \quad (1)$$

Here, $\Delta H$ and $H_r$ are the line widths and resonance fields corresponding to a particular frequency. $K_{sym}$ and $K_{asym}$ are the frequency dependent magnitudes of the absorptive and dispersive components present in the signals and $C$ is a constant offset. To understand the line shape, we look at the $K_{sym}/K_{asym}$ ratio for $R$ and $X$ as obtained from the DPPH sample. On fitting $R$ at 2 GHz it was found $|K_{sym}/K_{asym}| = 4.5658$, indicating $R$ is dominated by the symmetric component while for $X$, $|K_{sym}/K_{asym}| = 9.6943 \times 10^{-4}$ indicating the line shape is dominated by the dispersive component. In conventional EPR spectrometers, the derivative of the power absorbed is usually measured and the line shape is fitted to the Dysonian equation given by Eq. 2:[29]

$$\frac{dP}{dH_{dc}} \propto \frac{d}{dH_{dc}}\left[\left(\frac{\Delta H}{(H_{dc} - H_r)^2 + \Delta H^2}\right) + \left(\frac{\alpha(H_{dc} - H_r)}{(H_{dc} - H_r)^2 + \Delta H^2}\right)\right] \qquad (2)$$

The first term in the above equation describes the absorption while the second term represents the dispersion. $\alpha$ denotes the dispersion-to-absorption ratio. The asymmetricity is prominent in conducting samples since the electric and magnetic *rf* components in conducting samples become out of phase with each other leading to an admixture of the dispersion into the absorption spectra. $\alpha = 0$ when the skin effect is negligible as in insulating samples while $\alpha = 1$ for highly conducting samples where the skin depth is very small compared to the sample size. In this case the absorption and dispersion are of equal strength. So, a *dP/dH* measurement alone cannot isolate the absorption and dispersion effect whereas, the $R$ and $X$ responses from the magnetoimpedance measurements can provide this information and enable accurate analysis of the physical parameters.

In Fig. 2(a), the field dependence of $R$ for various frequencies from 1.5 GHz to 2.2 GHz are shown. The peak in $R$ shifts towards a higher magnetic field with increasing frequency of current. We performed the line shape analysis for all the frequencies and extracted the frequency dependent line widths ($\Delta H$) as well as the resonance fields ($H_r$) using Eq. 1. It is known that the resonance frequency ($f_r$) for EPR is proportional to the *dc* magnetic field and follows the relation $f_r = \left(\frac{\gamma}{2\pi}\right) H_{dc}$ where $\gamma$ is the gyromagnetic ratio ($\gamma = g\mu_B/\hbar$, where $g$ is the Landé g factor, $\mu_B$ is the Bohr magneton and $\hbar$ is the reduced Plank's constant). Therefore, with increasing $H_{dc}$ the resonance frequency increases linearly. This linear behavior was observed in the plot of $f_r$ vs $H_{dc}$ presented in Fig. 2(b) and we obtain $\gamma/2\pi = 2.799 \pm 0.0276$ MHz/Oe. This $\gamma/2\pi$ value corresponds to a Landé g value of 1.999 which is very close to the reported value of 2.0036. The line width in this frequency range was about 2 Oe which is consistent with the dilute nature of paramagnetic species (free radicals) in DPPH[30].

To verify the results obtained through the MI method, we measured the EPR spectra with a broad band ferromagnetic resonance spectrometer (Cryo-FMR by NanoOsc[TM] from Quantum Design Inc.USA). This spectrometer makes use of the lock-in technique and records the derivative of power absorbed (*dP/dH*) by the DPPH sample placed on top of a wave guide while $H_{dc}$ is swept for fixed *rf* excitations of 2 GHz, 4 GHz, 10 GHz and 12 GHz as shown in Fig. 3 (a) . We can see that the resonance field ($H_r$) which corresponds to the zero crossing point and amplitude of *dP/dH* increase with increasing frequency. The inset in Fig. 3(b) shows $H_r$ increasing linearly with frequency and $\gamma/2\pi = 2.801$ GHz/kOe, which is close to the value observed in the MI measurement. The *dP/dH* line shape was fitted to Eq. 4*:*

$$\frac{dP}{dH} = A_{asym}\frac{4\Delta H(H-H_r)}{[4(H-H_r)^2+(\Delta H)^2]^2} - A_{sym}\frac{(\Delta H)^2 - 4(H-H_r)^2}{[4(H-H_r)^2+(\Delta H)^2]^2} + C \quad (3)$$

where, $A_{asym}$ and $A_{sym}$ are the frequency dependent magnitudes of the absorptive and dispersive components present in the *dP/dH* signal. The intensity of absorption by the sample is directly proportional to the relative numbers of unpaired electrons in the sample. Therefore, a double integration of the derivative spectrum of absorbance can be used to estimate the spin concentration which can be utilized for quantitative EPR studies. In Fig. 3(b), the line shape fit, the single integration and the double integration of the *dP/dH* signal is presented.

In Fig. 4 (a) *R* is presented for various angles which $H_{dc}$ makes with $h_{rf}$. When $H_{dc}$ is perpendicular to $h_{rf}$ the signal is the most intense while it disappears when $H_{dc}$ is parallel to $h_{rf}$. The *R* response for different masses of DPPH is also presented in Fig. 4 (b). The signal strength is proportional to the mass of the DPPH with 23mg of DPPH exhibiting the largest response. Since *R* is proportional to $\chi''$, a single integration of the *R* response can be used to estimate the number of spins in the sample. The single integration of *R* is presented in the inset of Fig. 4(b) which provides the EPR intensity of absorption by the DPPH samples and increases with increase in number of spins. For 13 mg, 19 mg and 23mg of DPPH, we obtain 1.9853 x $10^{19}$, 2.9017 x $10^{19}$ and 3.5125 x $10^{19}$ spins, respectively.

**Conclusions**

In summary, we have presented a simple technique to probe EPR in the standard DPPH sample by passing high frequency currents in a copper stripcoil which surrounds the sample while measuring the magnetic field dependence of electrical impedance. We analyzed the line shapes using symmetric and asymmetric Lorentzian functions and the g-value was extracted. The electrical detection of EPR signal using an impedance analyzer which is traditionally used to characterize dielectric samples could aid the broad scientific community to probe EPR and understand the spin dynamics in the low frequency regime of the microwave spectrum.

**Conflicts of interest**

The authors declare no conflict of interest.

**Acknowledgements**

R.M. acknowledges the Ministry of Education, Singapore (grant number R144-000-381-112).

**References**

[1] E. K. Zavoisky, Zhur. Eksperiment. i Theoret. Fiz., Vol.15,1945, pp.344–350 For historical perspective, see G R Eaton, S. S Eaton and K. M Salikhov, *Foundations of modern EPR*, World Scientific, Singapore, 1998, pp.45–50.

[2] M.M Roessler, and E. Salvadori, Chem. Soc. Rev., 2018, **47**, 2534.

[3] K. L. Bower, Rev. Sci. Instr. 1977, **48**, 135; P. M. Lenhan, J. F. Conley Jr. J. Vac. Sci. Tech. B 1998, **16**, 2134. F. Jelezko, and J. Wrachtrup. Physica Status Solidi (a), 2006, **203**, 3207.

[4] K. A. Muller, W. Berlinger, and F. Waldner, Phys. Rev. Lett. 1968, **21**, 814

[5] K. L. Brower and S. M. Myers, Appl. Phys. Lett. (1990) **57**, 162.

[6] V. V. Khramstov, A. A. Bobkov, M. Tseytlin, and B. Driesschaert, Anal. Chem. (2017) **89**, 4758.

[7] S. Van Doorslaer and D.M. Murphy, *EPR Spectroscopy: Applications in chemistry and biology*, Ed. M. Drescher and G. Jeschke, Springer-verlag, Berlin 2012, pp. 1-29 .

[8] M. Che and E. Giamello, *Catalyst Characterization*, Ed. B. Imelik and J. C. Vedrine, Springer, Berlin 1994, pp. 131.

[9] Reijerse, E. and Savitsky, A. (2017). *Electron Paramagnetic Resonance Instrumentation* in eMagRes Ed. R. K. Harris and R. L. Wasylishen. doi:[10.1002/9780470034590.emrstm1511](10.1002/9780470034590.emrstm1511)

[10] Eaton, Gareth R. *EPR Imaging and In-vivo EPR*, CRC press, London 2018

[11] C. Clauss, M. Dressel, and M. Scheffler, J. Phys: Conference Series, 2015, **592**, 012146.

[12] Y. Wiemann, J. Simmendinger, C. Clauss, L. Bogani, D. Bothner, D. Koelle, R. Kleiner, M. Dressel, and M. Scheffler, Appl. Phys. Lett. 2015, **106**, 193505.

[13] K. Jing, Z. Lan, Z. Shi, S. Mu, X. Qin, X. Rong, and J. Du, Rev. Sci. Instrum 2019, **90**, 125109.

[14] H. Malissa, D. I. Schuster, A. M. Tryshkin, A. A. Houck, and S. A. Lyon, Rev. Sci. Instrum.2013, **84**, 025115.

[15] B. Johansson, S. Haraldson, L. Pettersson, and O. Beckman, Rev. Sci. Instrum., 1974, **45**, 1445.

[16] E. D. Dahlberg and S. A. Dodds, Rev. Sci. Instrum. 1981, **52**, 472.

[17] A. C. Torrezan, T. P. M. Alegre, and G. Medeiros-Ribeiro, Rev. Sci. Instrum. 2009, **80**, 075111.

[18] A. Ghirri, Bonizzoni, M. Righi, F. Fedele, G. Timco, R. Winpenny, and M. Affronte, Appl. Magn. Reson. 2015, **46**, 749.

[19] Misra, Sushil K., *Multifrequency electron paramagnetic resonance: theory and applications*. John Wiley & Sons, 2011.

[20] G. R. Eaton and and S. R. Eaton, *EPR: Instrumental methods*. Ed. L. J. Berliner and C. J. Bender, Springer, New York, 2004, pp.59-114.

[21] H. Ueda, Z. Kuri, and S. Shida, J. Chem. Phys., 1962, **36**, 1676.

[22] L. V. Panina, K. Mohri, K. Bushida, and M. Noda, J. Appl. Phys. 1994, **76**, 6198.

[23] U. Chaudhuri and R. Mahendiran, Appl. Phys. Lett. 2019 **115**, 092405.

[24] U. Chaudhuri, and R. Mahendiran, J. Magn. Magn. Mater. 2018, **488**, 165377.

[25] A. G. Marshall, and D. C. Roe, Anal. Chem. 1978, **50**, 756.

[26] P. Fajer, and D. Marsh, J. Magn. Reson. 1983, **55**, 205.

[27] C. Mailer, H. Thomann, B. H. Robinson, and L. R. Dalton, Rev. Sci. Instrum. 1980 **51**, 1714.

[28] J. S. Hyde, W. Froncisz, and A. Kusumi, Rev. Sci. Instr. 1982, **53**, 1934.

[29] V.A. Ivanshin, J. Deisenhofer, H-A. Krug Von Nidda, A. Loidl, A. A. Mukhin, A. M. Balbashov, and M. V. Eremin, Phys. Rev. B., 2000, **61**, 6213.

[30] W. R Hagen, J. Phys. Chem. A. 2019, **123,** 6986.

**Figure and Figure Captions**

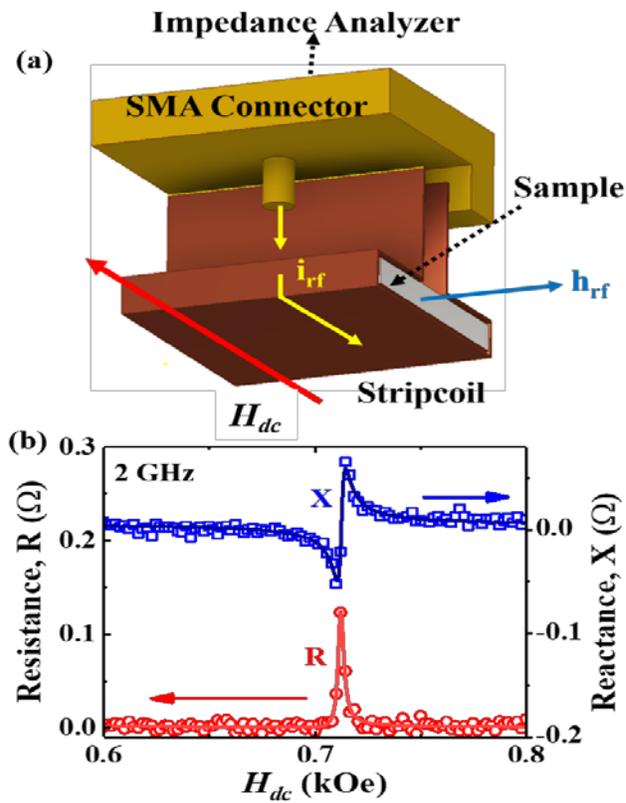

**Fig. 1.** (a) A schematic diagram of the copper stripcoil soldered to the SMA connector with $h_{rf}$ perpendicular to $H_{dc}$. (b) Resistance ($R$) and reactance ($X$) of the sample as measured by the stripcoil at room temperature with excitation frequency $f=2$ GHz.

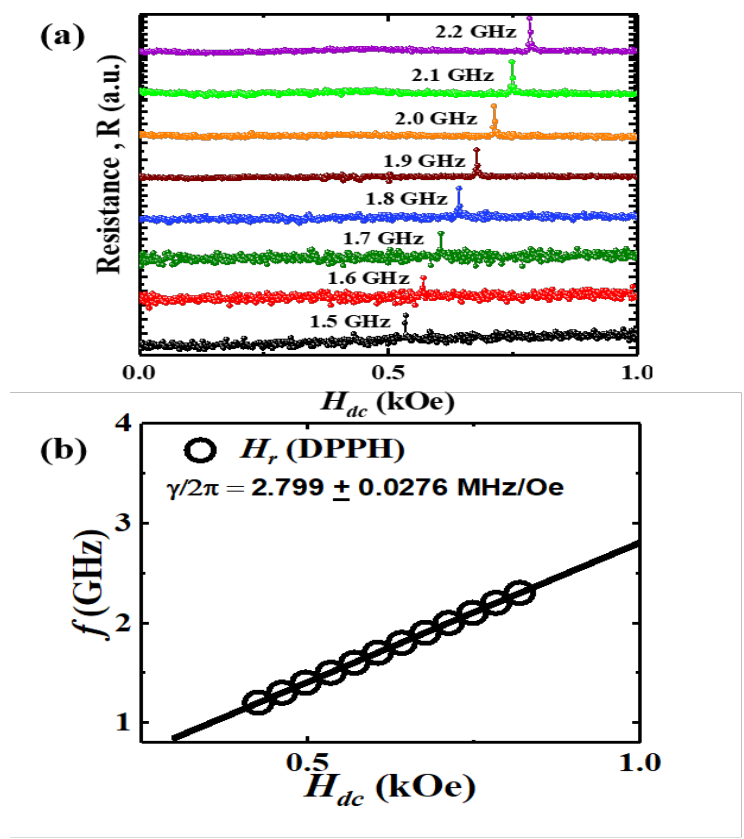

**Fig. 2.** (a) Resistance ($R$) of the copper stripcoil enclosing the DPPH sample as a function of $H_{dc}$ for different frequencies ($f$) of current in the stripcoil. (b) Plot of $f$ vs $H_{dc}$ with open circles used to depict the resonance fields ($H_r$) and the solid line illustrating the linear relationship indicating EPR.

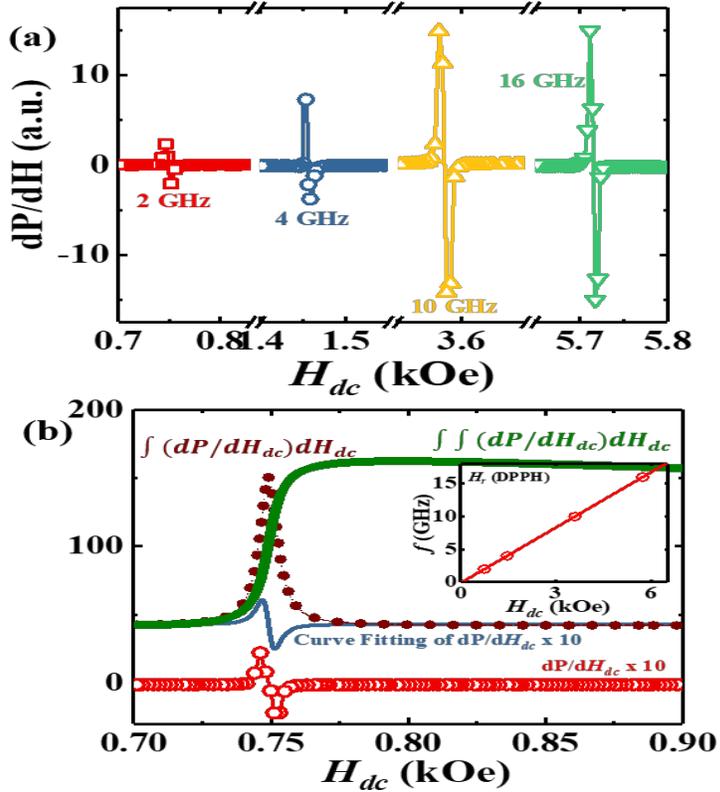

**Fig. 3.** (a) The EPR spectroscopic signal (*dP/dH*) for the DPPH sample measured using the Cryo-FMR spectrometer for excitation frequencies of 2 GHz (squares), 4 GHz (circles), 10 GHz (upward triangle) and 16 GHz (downward triangle). (b) *dP/dH* curve at 2 GHz along with the line shape fit using Eq. 4, the single integrated signal: $\int (dP/dH) dH_{dc}$ and the double integrated signal: $\int \int (dP/dH) dH_{dc}$ of the curve fitted *dP/dH* curve. Insert: Plot of *f* vs $H_{dc}$ with open circles used to depict the resonance fields ($H_r$) obtained using the *dP/dH* data.

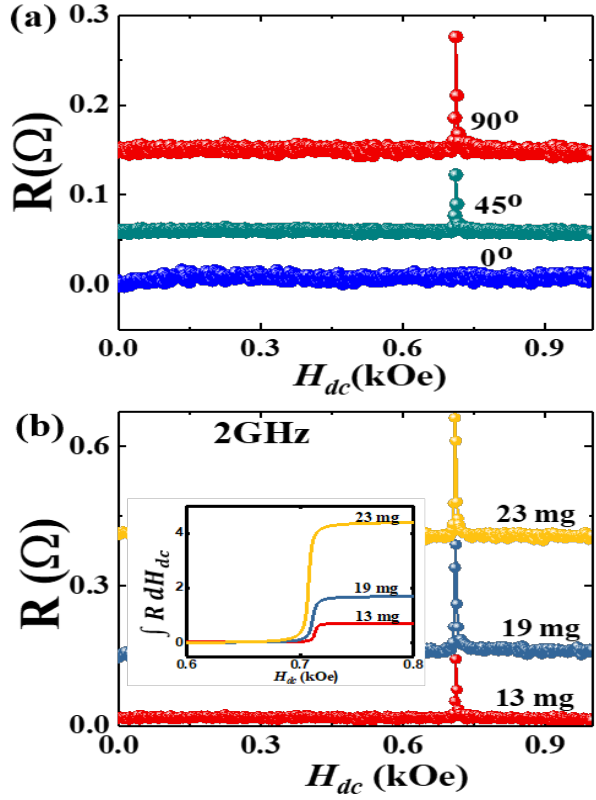

**Fig. 4.** (a) Resistance (*R*) of the copper stripcoil enclosing the DPPH when $h_{rf}$ makes 90°, 45° and 0° with $H_{dc}$. (b) Variation of *R* with $H_{dc}$ at 2 GHz for 23mg, 19mg and 13mg of DPPH. Insert: Single integration of the *R* curve for different masses of DPPH.